\documentstyle[12pt,epsfig,aaspp4]{article}
\begin{document}
\title{Could the Unusual Optical Afterglow of GRB 000301c Arise from
                  a Non-relativistic Shock with Energy Injection?}
\author{Z. G. Dai and T. Lu}
\affil{Department of Astronomy, Nanjing University, Nanjing 210093, China \\
E-mail: daizigao@public1.ptt.js.cn; tlu@netra.nju.edu.cn}

\begin{abstract}
Recent observations on the GRB 000301c afterglow show that three breaks 
appear in the R-band light curve, and in particular the decay slope at late 
times is as steep as $-3.0$. This unusual afterglow is clearly inconsistent 
with the standard afterglow shock model. Here we propose a non-standard 
model for the unusual R-band afterglow of GRB 000301c. In this 
model, an ultra-relativistic shock in a dense medium (``dirty environment")
rapidly evolved to the non-relativistic phase in initial 1 day. During such a phase, 
the shock happened to be caught up with by two energetic shells ejected from the 
central engine at two different times, and the shock was refreshed, leading to two 
flattenings of the light curve. After each interaction between the shock and shell, 
the afterglow decayed as $\propto t^{-3.0}$ if the electron distribution index 
of the shocked medium, $p\approx 3.4$, derived from the optical spectrum. 
Therefore, this model can provide an excellent explanation for the flattening and 
steepening features of the GRB 000301c optical afterglow light curve.
We further point out that the energy injection shells ejected from the central 
engine at later times may be material shells (e.g., in the massive star progenitor 
models related to black holes) or radiation shells (e.g., in the millisecond 
pulsar progenitor models).

\end{abstract} 


\section{Introduction}

The standard model of gamma-ray burst (GRB) afterglows assumes 
that a relativistic fireball is decelerating due to interaction with
the surrounding medium (for a review see Piran 1999). During such a 
deceleration, a relativistic forward shock forms and then produces 
an afterglow by synchrotron radiation and/or inverse Compton scattering. 
The simplest case of this model is that the surrounding medium is 
a homogeneous interstellar one with typical density of 
$\sim 1\,{\rm cm}^{-3}$. In this case, an optical afterglow light 
curve (e.g., GRB 970508) can be well fitted by a single power law until 
several months. However, this property, as we will see below, is clearly   
inconsistent with the peculiar optical afterglow of GRB 000301c.

GRB 000301c was independently detected by the All-Sky Monitor on the 
Rossi X-Ray Timing Explorer and by Ulysses and NEAR of the current 
Interplanetary Network on 2000 March 1.4108. The burst itself had a single 
peak lasting approximately 10 seconds  (Smith, Hurley \& Cline 2000). 
Its R-band afterglow on March 2.906 UT was first detected by UPSO 
(Masetti et al. 2000). This burst's redshift was measured as $z=2.0335
\pm 0.0003$ (Castro et al. 2000) by identifying weak metal lines 
in the afterglow's optical spectrum. According to the published papers
(Rhoads \& Fruchter 2000; Masetti et al. 2000; Sagar et al. 2000),
we can see the following features of the optical afterglow:  the 
R-band afterglow light curve in about 4 days after the burst was fitted 
approximately by a power-law with an index of $\alpha_1=-0.82\pm 0.20$,
and in later one day steepened based on another power law   
with an index of $\alpha_2=-3.0\pm 0.10$. However, during a period 
between the fifth and seventh days after the burst, the light curve 
flattened with the third temporal decay index of $\alpha_3=-0.53\pm 0.50$,
and subsequently till March 14.60 UT (the lastly observed time), 
steepened again based on a decay index $\alpha_4$ similar to 
$\alpha_2$. In addition, the V-band  and B-band afterglow could 
fade down almost simultaneously with the R-band one.

A successful scenario must explain the flattening and steepening features 
of the optical afterglow light curve of GRB 000301c. To our knowledge, 
four mechanisms have been proposed to account for steepening.  First, 
as the emission comes from slow-cooling electrons to fast-cooling 
electrons accelerated behind a relativistic shock in a homogeneous 
medium, its decay index steepens by a factor of 0.25 (Sari, Piran \& 
Narayan 1998), which is clearly inconsistent with the 
observational result. Second, as analyzed by many authors
(Vietri 1997; Dai \& Lu 1998a; M\'esz\'aros, Rees \& Wijers 1998; 
Panaitescu, M\'esz\'aros \& Rees 1998; Chevalier \& Li 1999, 2000), 
the afterglow from a relativistic shock in the wind medium must 
decay more rapidly than in the interstellar medium (ISM). For an adiabatic 
relativistic shock in the wind case, an electron distribution index of $p\sim 4.3$ 
is required by a large decay index of the late-time afterglow of GRB 000301c, 
$\alpha_2\approx\alpha_4\sim -3.0$. This would lead to a spectral index of 
$\beta\sim -1.7$, which is steeper than the observed one, $\beta_{\rm obs}=
-1.1\pm 0.1$, derived from the spectrum taken on 2000 March 3.47 UT  by 
Feng, Wang \& Wheeler (2000) and on March 14.61 UT by Sagar et al. (2000), 
respectively. Third, the steepening of a late-time optical afterglow light curve 
may be caused by lateral spreading of a jet (Rhoads 1999; Sari, Piran \& 
Halpern 1999). A difficulty for this mechanism is that the degree of steepening 
found by numerical studies (e.g., Panaitescu \& M\'esz\'aros 1999; Moderski, 
Sikora \& Bulik 2000; Huang, Dai \& Lu 2000; Wei \& Lu 2000) when two 
effects such as the equal-time surface and detailed dynamics of the jet 
are considered is much weaker than the one analytically predicted.
Finally, we recently suggested that the evolution of a relativistic shock 
in a dense medium to the non-relativistic phase should lead to steepening 
of an afterglow light curve (Dai \& Lu 1999). We found that this model is 
quite consistent with the observations on the GRB 990123 afterglow 
if the medium density is about $10^6\,{\rm cm}^{-3}$. Furthermore, 
as shown analytically and numerically by Dai \& Lu (2000a) and 
Wang, Dai \& Lu (2000), this model can also well fit all the GRB 980519 
afterglow data.

Energy injection from the GRB central engine to its postburst 
shock has  been widely argued to be a plausible scenario causing 
flattening of an afterglow light curve. This scenario can be realized by 
two different mechanisms: (1) The central engine may eject some shells 
with different Lorentz factors at different times. As  the outer
shells move outward, they begin to interact with the surrouding medium 
and decelerate, forming a forward shock (blast wave). Eventually the slower 
inner shells catch up with the outer shells. The interaction of slow shells with 
faster shells that have been slowed down implies refreshment of  the shock, 
leading to a flattening of the afterglow light curve (Rees \& M\'esz\'aros 1998; 
Panaitescu et al. 1998; Kumar \& Piran 2000; Sari \& M\'esz\'aros 2000). 
(2) If the GRB central engine is a strongly magnetized millisecond pulsar,
its rotational energy input to the postburst shock through magnetic dipole
radiation also results in a flattening of the afterglow light curve (Dai \& Lu
1998b, 1998c, 2000a). In this Letter we argue that  combination of our 
dense medium model with such an energy injection scenario can provide
an excellent explanation for the peculiar optical afterglow light curve of 
GRB 000301c.  

\section{The Model}

The dense (``dirty") environment of GRBs has been discussed in the literature. 
For example, collisions of relativistic nucleons with a dense cloud is 
suggested by Katz (1994) to explain the delayed hard photons from 
GRB 940217. The presence of an iron emission line in the X-ray afterglow 
spectrum of GRB 970508 and GRB 970828 reported by Piro et al. (1999) 
and Yoshida et al. (1999) requires that the ambient medium of these bursts 
is rather dense (Lazzati, Campana \& Ghisellini 1999). The steepening of 
the light curves of some optical afterglows (e.g., GRB 990123 and 
GRB 980519) may be due to the transition to the non-relativistic phase. 
This also requires that the medium density is as high as $10^6\,
{\rm cm}^{-3}$ (Dai \& Lu 1999, 2000a). The medium with a similar density 
is invoked by Dermer \& B\"ottcher (2000) to resolve the ``line-of-death" 
objection to the GRB synchrotron shock model. This work is guided by 
the optical observations of $\eta$ Carinae (a best-studied massive star),
whose environment is a dense cloud (Davidson \& Humphrey 1997). 
In addition, dense media may appear in some energy source models, 
e.g., failed supernovae (Woosley 1993), hypernovae (Paczy\'nski 1998), 
supranovae (Vietri \& Stella 1998), phase transitions of neutron stars 
to strange stars (Dai \& Lu 1998b; Wang et al. 2000),   and anisotropic 
supernovae (Wheeler et al. 2000).   

Based on these motivations, we here assume that the surrounding 
medium is dense and perhaps at different times the central engine 
ejects several shells, some of which are relativistic and others 
are non-relativistic. It is widely believed that a slow shell 
contains more energy than a faster shell which is ejected at an earlier time. 
Collisions between shells with large Lorentz factors give rise to internal shocks 
which are expected to produce GRBs. After then, these merging shells decelerate, 
leading to a forward shock (blast wave), as they sweep up the dense medium. 
The shock can be thought to be adiabatic unless the electron energy fraction 
of the shocked medium is as large as $\sim 1$ (Dai, Huang \& Lu 1999). 
The Blandford-McKee (1976) self-similar solution gives the Lorentz factor 
of an adiabatic relativistic shock:  $\gamma = 1.0E_{52}^{1/8}n_5^{-1/8}
t_{\rm day}^{-3/8}[(1+z)/2]^{3/8}$, where $E_{52}\times 10^{52}
{\rm ergs}$ is the total isotropic energy, $n=n_5\times 10^5\,{\rm cm}^{-3}$ 
is the medium density, $t_{\rm day}=t/1\,{\rm day}$ is the observer time, 
and $z$ is the the redshift of the source. This equation implies that, at time 
\begin{equation}
t_{\rm nr}=0.7E_{52}^{1/3}\left(\frac{n}{10^6{\rm cm}^{-3}}\right)^{-1/3}
\left(\frac{1+z}{3}\right)\,{\rm  days}, 
\end{equation}
the shock begins to enter the non-relativistic phase. 

According to equation (1), therefore, we find that if GRB 000301c was 
in a dense medium with density of $\sim 10^6\,{\rm cm}^{-3}$, 
its postburst shock would be non-relativistic at a time less 
than 1 day after the burst. If, during such a period, this shock happened to be 
caught up with by an energetic homogeneous shell which had been ejected 
from the central source at some time, then the shock would be refreshed 
and its energy evolved based on 
\begin{equation}
E_{\rm shock}=E_0+\left(\frac{1}{1+z}\right)\int_{t_0}^tL(t)dt,
\end{equation}
where $E_0$ is the initial energy of the shock, $t_0$ is the time at which
the shell started to inject energy, and $L(t)$ is the injection luminosity.
Assuming that the shell's velocity, energy and width are $v_{\rm shell}$, 
$E_{\rm shell}$ and $\Delta R$, and the shock's velocity is $v_{\rm shock}$,
we can write the injection luminosity approximately as
\begin{equation}
L(t)\approx \frac{v_{\rm shell}-v_{\rm shock}}{\Delta R}E_{\rm shell}
\approx \frac{v_{\rm shell}}{\Delta R}E_{\rm shell}.
\end{equation}
In writing the second expression, we have assumed $v_{\rm shell}\gg 
v_{\rm shock}$. The energy of the non-relativistic shocked medium can be 
approximated by 
\begin{equation}
E_{\rm shock}\approx\frac{2\pi}{3}v_{\rm shock}^2
R_{\rm shock}^3nm_p\propto v_{\rm shock}^2R_{\rm shock}^3,
\end{equation}
where $R_{\rm shock}$ is the shock's radius, 
$m_p$ is the proton mass and $c$ is the speed of light. Assuming that 
$E_{\rm shell}\gg E_0$ and the energy which the shock had obtained from 
the shell is much larger than $E_0$ when $t\gg t_0$, combination of 
equation (2) with equations (3) and (4) leads to 
\begin{equation}
v_{\rm shock}^2R_{\rm shock}^3\propto t.
\end{equation}
Because $R_{\rm shock}\propto v_{\rm shock}t$, we easily find 
\begin{equation}
v_{\rm shock}\propto t^{-2/5}.
\end{equation}

In the following we consider only synchrotron radiation from the shock and 
ignore synchrotron self absorption. To analyze the spectrum and light curve, 
one needs to know two crucial frequencies: the synchrotron peak frequency 
($\nu_m$) and the cooling frequency ($\nu_c$). Unfortunately, these 
frequencies are dependent on two unknown parameters: the electron energy 
fraction ($\epsilon_e$) and the magnetic energy fraction ($\epsilon_B$) 
of the shocked medium. Even so, the optical-band frequency is usually much 
higher than the $\nu_m$ of a late-time afterglow. From equation (6), 
we find the shock's radius $R_{\rm shock}\approx (5/3)v_{\rm shock}t
/(1+z)\propto t^{3/5}$ and the internal field strength $B=(4\pi \epsilon_B
nm_pv_{\rm shock}^2)^{1/2}\propto t^{-2/5}$.  The typical electron 
Lorentz factor $\gamma_m\approx [m_p/(2m_e)]\epsilon_e
(v_{\rm shock}/c)^2\propto t^{-4/5}$ and the synchrotron peak frequency 
$\nu_m=\gamma_m^2(eB)/[(1+z)2\pi m_ec]\propto t^{-2}$. The cooling 
Lorentz factor $\gamma_c=6\pi m_ec(1+z)/(\sigma_TB^2t)\propto 
B^{-2}t^{-1}$ with $\sigma_T$ being the Thomson scattering cross 
section (Sari, Piran \& Narayan 1998) and the cooling frequency 
$\nu_c=\gamma_c^2(eB)/[(1+z)2\pi m_ec]\propto B^{-3}t^{-2}\propto 
t^{-4/5}$. The synchrotron peak flux decays as $F_{\nu_m}=
(1+z)N_eP_{\nu_m}/(4\pi D_L^2)\propto R^3B\propto t^{7/5}$, 
where $N_e=(4\pi/3)R^3n$ is the total number of swept-up 
electrons in the postshock fluid, $P_{\nu_m}=m_ec^2\sigma_T B/(3e)$ 
is the power radiated per electron per unit frequency and $D_L$ is the 
luminosiy distance from the source. According to these scaling laws, 
we further derive the spectrum and light curve of the afterglow 
\begin{equation}
F_\nu=\left \{
       \begin{array}{ll}
         (\nu/\nu_m)^{-(p-1)/2}F_{\nu_m} \propto \nu^{-(p-1)/2}
         t^{(12-5p)/5} & {\rm if}\,\,\,\,  \nu \le \nu_c \\
         (\nu_c/\nu_m)^{-(p-1)/2}(\nu/\nu_c)^{-p/2}F_{\nu_m}\propto
              \nu^{-p/2}t^{2-p} & {\rm if}\,\,\,\, \nu>\nu_c,
        \end{array}
       \right.
\end{equation}
where $p$ is the electron distribution index 
(Dai \& Lu 2000a). We note that if $p=3.4$, then
$\alpha=(12-5p)/5=-1.0$ and $\beta=-(p-1)/2=-1.2$ are consistent with the 
GRB 000301c R-band afterglow data both in initial 4 days and during 
a period between the fifth and seventh days after the burst.  These data 
indicate $\alpha_1=-0.82\pm 0.20$, $\alpha_3=-0.53\pm 0.50$ and 
$\beta_{\rm obs}=-1.1\pm 0.1$, which imply $\alpha_{\rm obs}\approx 
5\beta_{\rm obs}/6$. If the afterglow were radiated by fast-cooling 
electrons in the shocked medium, we would find $\alpha=2(1-\beta)$, 
which is clearly inconsistent with the observational result. Therefore, 
the GRB 000301c R-band afterglow arose from those slow-cooling 
electrons in the shocked medium.

After the shell had input all its energy to the shock, from equations (2)
and (4), the shock's velocity evolved as $v_{\rm shock}\propto t^{-3/5}$ 
and thus the spectrum and light curve of the afterglow became
\begin{equation}
  F_\nu\propto \left \{
      \begin{array}{ll}
           \nu^{-(p-1)/2}t^{(21-15p)/10} & {\rm if}\,\,\,\, \nu \le\nu_c \\
           \nu^{-p/2}t^{(4-3p)/2} & {\rm if}\,\,\,\, \nu>\nu_c,
        \end{array}
       \right.
\end{equation}
where the $\nu_c$ is different from that in equation (7) 
(Wijers, Rees \& M\'esz\'aros 1997; Dai \& Lu 1999, 2000a).
We can see from this equation that in the case of $p=3.4$, the model's decay 
index $\alpha=(21-15p)/10=-3.0$ is quite consistent with the observational 
data of the GRB 000301c R-band afterglow both during a period between 
the fourth and fifth days after the burst and at later times, 
$\alpha_2\approx \alpha_4=-3.0\pm 0.1$.

\section{Discussion}

We have shown that our dense medium model combined with the 
energy injection scenario can provide a plausible explanation for the unusual
optical afterglow of GRB 000301c. Now we want to discuss the two energy 
injection mechanisms in some details. In the first mechanism, the shells which 
input their energy into the shock are {\em  material shells}, whose energy 
source may be the core collapse of massive stars to black holes. 
As the shock decelerates, such a shell eventually catches up with the shock. 
During such an interaction, two additional shocks (a forward shock and 
a reverse shock) might form and perhaps give rise to some observational 
effects on afterglows. Kumar \& Piran (2000) analyzed such effects 
of ultra-relativistic shocks. We will make a detailed analysis for 
non-relativistic shocks and discuss their possible effects on afterglows 
in a future paper (Dai \& Lu 2000b). It is interesting to note that a small but 
discernible variability appears in the GRB 000301c optical afterglow light 
curve in initial 4 days, which could be due to these additional shocks.  

In the second mechanism, the shells which input their energy 
into the shock may arise from a strongly magnetized millisecond 
pulsar.  In this case, $t_0\approx 0$ in equation (2) and 
$v_{\rm shell}=c$ in equation (3). Because the magnetic dipole radiation 
luminosity $L(t)\propto (1+t/T)^{-2}$, where $T$ is the characteristic 
spin-down age, $L(t)$ can be thought of as a constant for $t<T$ while 
$L(t)$ decays as $\propto t^{-2}$ for $t\gg T$. In addition, we define
a timescale $t_{\rm cr}$ based on $t_{\rm cr}=E_0/L$. As found by Dai 
\& Lu (1998b, 1998c), the pulsar energy input effect can be neglected for
both $t<t_{\rm cr}$ and $t>T$, but the pulsar inputs its rotational energy
into the shock  at an approximately constant rate for $t_{\rm cr}<t<T$. 
This further implies $\Delta R\approx c(T-t_{\rm cr})$ in equation (3).  
Therefore, the energy injection shells in this case are 
{\em radiation shells}. It should be pointed out that the GRB source models 
involving strongly magnetized millisecond pulsars include accretion-induced
collapses of magnetized white dwarfs (Usov 1992; Blackman, Yi \& Field 
1996; Ruderman, Tao, \& Klu\'zniak 2000), mergers of two neutron stars 
if the equation of state for neutron matter is moderately stiff to stiff 
(Klu\'zniak \& Ruderman 1998), phase transitions of neutron 
stars (Dai \& Lu 1998b), R-mode-induced explosions in low-mass
X-ray binaries (Spruit 1999),  and anisotropic supernovae 
(Wheeler et al. 2000). All these models have been proposed to 
possibly produce GRBs with long durations, which are consistent 
with GRB 000301c.    

There are two flattenings in the optical afterglow light 
curve of GRB 000301c. To interpret this, we require, 
in the first mechanism, only two energetic material shells which caught
up with the shock at two different times. In the second mechanism, how did 
the central pulsar input its rotational energy into the shock twice? We 
envision that a strongly magnetized, rapidly rotating newborn neutron star  
first spun down through magnetic dipole radiation and thus input most of
its rotational energy to the postburst shock, leading to the first flattening of the 
afterglow light curve. As the neutron star spun down to some extent, it would 
undergo a phase transition to become a more compact star (e.g., a strange star)
(Dai \& Lu 1998b; Wang et al. 2000). This case is somewhat similar to the one
discussed by Vietri \& Stella (1998), where as the neutron star spins down it 
will collapse to a black hole. Assuming that $J$ is the angular momentum at the
phase transition, and $I_{\rm NS}$ and $I_{\rm MCS}$ are the moments of 
inertia of the neutron star and the more compact star respectively, angular 
momentum conservation implies that the rotational energy of the post-transition 
star would significantly increase, that is, $E_{\rm MCS}=J^2/(2I_{\rm MCS})\gg
E_{\rm NS}=J^2/(2I_{\rm NS})$ because $I_{\rm MCS}$ may be much less
than $I_{\rm NS}$. The post-transition star would subsequently spin 
down through magnetic dipole radiation and inject its rotational energy to 
the postburst shock, possibly resulting in the second flattening of 
the afterglow light curve. 

The infrared (2.1$\mu$m) light curve of the GRB 000301c 
afterglow is different from the R-band, B-band and V-band light 
curves. The latter present the flattening and steepening features but 
the former shows a well-sampled break in the decay index at $t\approx 3.5$ 
days after the burst. The early time index at infrared frequency is very 
shallow ($\sim -0.1$), while the late time index is steep ($-2.2$) (Rhoads \& 
Fruchter 2000). We here argue that such infrared and R-band emissions 
might result from different radiation mechanisms. Otherwise, (1) their 
temporal decay should be almost independent of frequency, which 
is contrary to the observational result; (2) the ratio of their fluxes should 
approximately be $(\nu_{\rm K'}/\nu_{\rm R})^{-(p-1)/2}
\sim 3.7$ where $p=3.4$. However, this ratio derived from the observed data 
in about three days after the burst is about 10. Therefore, the infrared 
emission was unlikely to arise from the non-relativistic shock. In addition,
the singly-broken power-law decay of the infrared emission is reminiscent of 
a relativistic jetted shock. However, as analytically shown by Rhoads (1999) and 
Sari et al. (1999),  the time index of an afterglow from a lateral-spreading 
jet evolves from $\alpha=3(1-p)/4$ or $(2-3p)/4$ to $\alpha=-p$. 
The observed early-time index at infrared frequency ($\alpha\sim -0.1$) 
leads to $p\sim 1.1$ or $0.8$,  implying that the late-time index 
$\alpha\sim -1.1$ or $-0.8$, which is clearly inconsistent with the 
observations. Therefore, the infrared emission couldn't arise from a 
relativistic jetted shock. We note that for the GRB 990123 
afterglow its infrared light curve is much different from that at R-hand 
frequency (Kulkarni et al. 1999).  The observed infrared afterglow 
emission of GRB 000301c might be produced by dust sublimation 
(Waxman \& Draine 2000; Esin \& Blandford 2000). However, whether 
or not this possibility is correct needs further analytical and numerical studies.

\section{Conclusions} 

Many optical afterglows can be well fitted by a single power-law decay, which 
supports the standard relativistic shock model. But, three breaks appear in the 
R-band afterglow light curve of GRB 000301c, and in particular the decay index 
at late times is as steep as $-3.0$. This unusual afterglow is clearly inconsistent 
with the standard model. Following Dai \& Lu (1999, 2000a), we have here 
proposed a non-standard shock model for the unusual R-band afterglow of 
GRB 000301c. In this model, an ultra-relativistic shock in a dense medium 
(``dirty environment") rapidly evolved to the non-relativistic phase in initial 1 day. 
During such a transition, the shock was by chance caught up with by 
one energetic shell ejected from the central engine at a later time, 
and the shock was refreshed, leading to the first flattening of the 
light curve. Once the interaction between the shock and shell finished, 
the afterglow started to decay as $\propto t^{-3.0}$ if the electron 
distribution index of the shocked medium $p\approx 3.4$ derived from 
the optical spectrum. One day later, the shock was by chance caught up with 
by another more energetic shell, resulting in the second flattening of the light 
curve. After this interaction, the shock evolved based on the Sedov-Taylor self-similar 
solution without energy injection and the afterglow light curve steepened again.
The energy injection shells ejected from the central engine at later times may 
be material shells (e.g., in the massive star progenitor models related to black holes) 
or radiation shells (e.g., in the millisecond pulsar progenitor models).
  
\acknowledgments

We would like to thank Drs. Y. F. Huang, Y. P. Wang and D. M. Wei 
for helpful discussions. This work was supported partially by 
the National Natural Science Foundation of China (grants 19825109 
and 19773007) and partially by the National Project on Fundamental 
Researches.

\end{document}